\documentclass[prl,aps,twocolumn,amstex]{revtex4}
\usepackage{amsmath}
\usepackage{amssymb}
\usepackage{bm}
\usepackage{graphicx}
\newcommand{\vek}[1]{\bm{\mathrm{#1}}}
\newcommand{\Jz}{M}
\newcommand{\Lz}{M_L}
\newcommand{\Sz}{M_S}
\newcommand{\Clebsch}[2]{C_{#1}^{#2}}
\begin{document}

\title
{Spontaneous generation of spin-orbit coupling 
in magnetic dipolar Fermi gases}

\author{T. Sogo$^1$, M. Urban$^1$, P. Schuck$^{1,2}$, T. Miyakawa$^3$}
\affiliation{$^1$
Institut de Physique Nucl\'eaire, CNRS-IN2P3 
and Universit\'e Paris-Sud, 91406 Orsay Cedex, France}
\affiliation{$^2$
Laboratoire de Physique et Mod\'elisation 
des Milieux Condens\'es, CNRS and Universit\'e
Joseph Fourier, 25 Avenue des Martyrs, 
Bo\^ite Postale 166, F-38042 Grenoble Cedex 9, France}
\affiliation{$^3$
Faculty of Education, 
Aichi University of Education, Hirosawa 1, Igaya-cho, Kariya 448-8542, Japan}

\begin{abstract}
The stability of an unpolarized two-component dipolar Fermi gas is
studied within mean-field theory. Besides the known instability
towards spontaneous magnetization with Fermi sphere deformation,
another instability towards spontaneous formation of a spin-orbit
coupled phase with a Rashba-like spin texture is found. A phase
diagram is presented and consequences are briefly discussed.
\end{abstract}


\maketitle

Recently, Bose-Einstein condensation of Dysprosium 164 has been
achieved \cite{lu11}. 
$^{164}$Dy is an atom with magnetic dipole moment $10\mu_B$
($\mu_B$ being the Bohr magneton). Also a dipolar Fermi
gas was recently produced in the form of a gas of fermionic molecules
with large electric dipole moment \cite{Jin11}. This gas was, however,
not degenerate. But since there exists a fermionic Dy isotope,
$^{163}$Dy, we can expect in the near future the realization of a
degenerate Fermi gas of atoms with large magnetic moment.

Dipolar atoms interact via the dipole-dipole interaction (DDI), which
is anisotropic and long-ranged. In Fermi gases, this interaction has
the effect to deform the Fermi surface. This phenomenon has been
studied in Refs.~\cite{msp08,ff09,ff10}. In nuclear physics, the
so-called tensor force has a similar structure as the DDI
\cite{ringschuck}. The effect of the tensor force in nuclear matter
has been investigated in the framework of Landau Fermi liquid theory,
e.g., with regard to the spin susceptibility \cite{jerzak,pethick}.

Since their recent realization \cite{Lin11}, cold atomic systems with
artificial spin-orbit coupling (SOC), in two as well as three
dimensions, have received tremendous attention (see, e.g.,
\cite{Sau11,Yu11,Jiang11}). In this letter, we show that in three
dimensional unpolarized dipolar Fermi gases, the DDI can give rise to
an instability towards spontaneous formation of a phase with SOC. This
could be an alternative way, maybe simpler than the artificial SOC, to
produce SOC in ultracold Fermi gases, and opens wide perspectives
which have been intensely discussed in the very recent literature. For
instance, SOC may have important consequences for pairing
\cite{Yu11,Jiang11}.

The possibility of a spontaneous generation of a SOC phase will be
studied within the mean-field and random-phase approximation (RPA)
approach. We will first formulate the theory for a magnetic dipolar
Fermi gas with arbitrary spin $s$ and then specialize to $s=1/2$.

The magnetic DDI between two atoms with dipole moments $\vek{d}_1$ and
$\vek{d}_2$ is given by 
\begin{eqnarray}
V_{dd}(\vek{r})
=
-\frac{3}{r^3}
\Big(\frac{(\vek{r}\cdot \vek{d}_1)(\vek{r}\cdot\vek{d}_2)}{r^2}
-\frac{\vek{d}_1\cdot \vek{d}_2}{3}\Big)\,,
\label{eq-realspaceDDI}
\end{eqnarray}
where 
$\vek{r}$ is the distance between the atoms.

Let us consider a uniform gas of fermionic atoms having a magnetic
dipole moment $\vek{d}=d_0 \vek{s}$, where $\vek{s}$ is the spin
operator and $d_0$ characterizes the magnitude of the dipole
moment. The Hamiltonian is given by
(we use units with $\hbar=1$)
\begin{multline}
H = \sum_{\alpha}\int \frac{d^3k}{(2\pi)^3}
\frac{k^2}{2m}
c^\dagger_{\vek{k},\alpha}c_{\vek{k},\alpha}
\\
+\frac{1}{2}
\sum_{\alpha\beta\alpha'\beta'}
\int \frac{d^3k}{(2\pi)^3}\frac{d^3k'}{(2\pi)^3}
\frac{d^3q}{(2\pi)^3}V_{\alpha\beta,\alpha'\beta'}(\vek q)
\\
\times c^\dagger_{\vek{k}+\vek{q},\alpha}c^\dagger_{\vek{k}'-\vek{q},\beta}
c_{\vek{k}',\beta'}c_{\vek{k},\alpha'}\,,
\end{multline}
The interaction includes the contact interaction as well
as the dipole-dipole interaction:
$V_{\alpha\beta,\alpha'\beta'}(\vek q)
=
V^{c}_{\alpha\beta,\alpha'\beta'}
+
V^{dd}_{\alpha\beta,\alpha'\beta'}(\vek q)\,.
$

The contact interaction is written as
$V^{c}_{\alpha\beta,\alpha'\beta'} =
g\delta_{\alpha\alpha'}\delta_{\beta\beta'}$.  For simplicity we
assume it to be spin independent, although in general its strength
depends on the total spin of the two interacting atoms \cite{yh99}.
This approximation becomes exact in the case of $s=1/2$ atoms, which
we shall discuss below, since in this case only the spin-singlet
channel contributes.

The dipole-dipole interaction $V^{dd}(\vek{q})$ is the Fourier
transform of Eq.~(\ref{eq-realspaceDDI}),
\begin{equation}
V^{dd}_{\alpha\beta,\alpha'\beta'}(\vek q)
=
4\pi c_{dd}
\Big(\frac{(\vek{q}\cdot\vek{s}_{\alpha\alpha'})
(\vek{q}\cdot\vek{s}_{\beta\beta'})}{q^2}
-\frac{\vek{s}_{\alpha\alpha'}\cdot\vek{s}_{\beta\beta'}}{3}\Big)\,,
\label{eq-ddinteraction}
\end{equation}
where 
$c_{dd}=d_0^2$  denotes the
coupling constant of the dipole-dipole interaction and
$\vek{s}_{\alpha\alpha'}$ is the matrix element of the spin operator
between the basis spin functions $(s_m)_{\alpha\alpha'} =
\sqrt{s(s+1)}\Clebsch{s\alpha' 1 m}{s \alpha}$ with $s_{\pm1} = \mp (s_x\pm
is_y)/\sqrt{2}$, $s_0=s_z$, and the Clebsch-Gordan coefficient
$\Clebsch{s\alpha' 1 m}{s \alpha}$ in the notation of Ref.~\cite{vmk88}.

We assume for the moment that the ground state of the system is spin
symmetric. In this case, the DDI does not contribute to the mean
field, which implies that all spin components have the same spherical
Fermi surface (note that, as shown in \cite{ff09,ff10}, the Fermi
surface deforms if the ground state is spin asymmetric, i.e., if the
gas is fully or partially polarized). Then the occupation numbers are
given by $\rho_{\alpha\beta}(\vek{k}) =\langle
c^\dagger_{\vek{k},\beta} c_{\vek{k},\alpha}\rangle
=\delta_{\alpha\beta}\rho(\vek{k})
=\delta_{\alpha\beta}\theta(k_F-k)$, where $k_F$ denotes the Fermi
momentum. The single-particle energies are $\varepsilon_{\vek{k}}
=k^2/(2m)+2sgn$, where $2sgn$ is the Hartree-Fock mean field, $n =
k_F^3/(6\pi^2)$ being the density per spin state.

We will use RPA theory in order to investigate for which
parameters the symmetric ground state gets unstable with respect to
the formation of more interesting asymmetric phases. To that end, we
calculate the spectrum of the collective zero-sound modes: A vanishing
or imaginary frequency indicates an instability. Let us consider the
retarded response function
\begin{multline}
i\Pi_{\alpha\beta,\alpha'\beta'}(\vek{k},\vek{k}',\vek{q},t-t') = \theta(t-t')
\\
\times \langle 
[c^\dagger_{\vek{k},\beta}(t)
c_{\vek{k}+\vek{q},\alpha}(t),
c^\dagger_{\vek{k}'+\vek{q},\alpha'}(t')
c_{\vek{k}',\beta'}(t')]
\rangle\,.
\end{multline}
Within RPA, it is obtained as the solution of the integral equation
\begin{multline}
\Pi_{\alpha\beta,\alpha'\beta'}(\vek{k},\vek{k}',\vek{q},\omega) = 
\Pi^0(\vek{k},\vek{q},\omega)
(2\pi)^3\delta(\vek{k}-\vek{k}')\delta_{\alpha\alpha'}\delta_{\beta\beta'}
\\
+
\Pi^0(\vek{k},\vek{q},\omega)
\sum_{\alpha_1\beta_1}
\int \frac{d^3k_1}{(2\pi)^3}
f_{\alpha\beta_1,\beta\alpha_1}(\vek{q},\vek{k}-\vek{k}_1)\\
\times
\Pi_{\alpha_1\beta_1,\alpha'\beta'}(\vek{k}_1,\vek{k}',\vek{q},\omega)\,,
\label{eq-RPAgeneral}
\end{multline}
where 
$f_{\alpha\beta,\alpha'\beta'}(\vek{q},\vek{k}-\vek{k}') =
V_{\alpha\beta,\alpha'\beta'}(\vek{q}) -
V_{\alpha\beta,\beta'\alpha'}(\vek{k}-\vek{k}')$ 
denotes the
antisymmetrized matrix element of the interaction and
$\Pi^0(\vek{k},\vek{q},\omega) = [\rho(\vek{k})-\rho(\vek{k}+\vek{q})]
/(\omega-\varepsilon_{\vek{q}+\vek{k}}+\varepsilon_{\vek{k}}+i\eta)$
is the non-interacting response function.

In the limit $q \ll k_F$, the response function is concentrated at $k
= k' = k_F$ and it depends only on the directions of $\vek{k}$ and
$\vek{k}'$ and on the dimensionless quantity $\tilde{\omega} =
m\omega/(k_Fq)$ (without loss of generality we suppose that $\vek{q} =
q\vek{e}_z$). As usual in Landau Fermi liquid theory, we expand the
angular dependence in spherical harmonics $Y_{L\Lz}$. A complication
arises from the fact that the DDI is not diagonal in orbital angular
momentum $L$ and spin $S$ of the excitation, but only in the total
angular momentum $J$. We introduce a multi-index $\Lambda = LSJ\Jz$
and define the angular momentum projected response function as
\begin{multline}
\tilde{\Pi}_{\Lambda\Lambda'}(\tilde{\omega})
=
\sum_{\Sz \Lz \Sz' \Lz'}
\Clebsch{L \Lz S \Sz}{J \Jz}
\Clebsch{L' \Lz' S' \Sz'}{J' \Jz'}\\
\times
\sum_{\alpha\beta\alpha'\beta'}
(-1)^{s-\beta}\Clebsch{s \alpha s\, -\beta}{S \Sz}
(-1)^{s-\beta'}\Clebsch{s \alpha' s\, -\beta'}{S' \Sz'}
\\
\times
\int \frac{d^3k}{(2\pi)^3}\frac{d^3k'}{(2\pi)^3}
Y^*_{L \Lz}(\Omega_k)
Y_{L' \Lz'}(\Omega_k')
\\
\times
\frac{(2\pi)^3}{mk_F}
\Pi_{\alpha\beta,\alpha'\beta'}(\vek{k},\vek{k}',\vek{q},\omega)\,.
\end{multline}
Then the RPA equation (\ref{eq-RPAgeneral}) reduces to a matrix
equation:
\begin{equation}
\tilde{\Pi}_{\Lambda\Lambda'}(\tilde{\omega})
=
\tilde{\Pi}^0_{\Lambda\Lambda'}(\tilde{\omega})
+
\sum_{\Lambda_1\Lambda_2}
\tilde{\Pi}^0_{\Lambda\Lambda_1}(\tilde{\omega})
F_{\Lambda_1\Lambda_2}
\tilde{\Pi}_{\Lambda_2\Lambda'}(\tilde{\omega})\,,
\label{eq-RPAeq-J}
\end{equation}
where
\begin{multline}
\tilde{\Pi}^0_{\Lambda\Lambda'}(\tilde{\omega})
=
\delta_{S S'}\delta_{\Jz \Jz'}
\sum_{\Lz\Sz} \Clebsch{L\Lz S \Sz}{J\Jz} \Clebsch{L'\Lz S \Sz}{J'\Jz}
\\ \times
\int d\Omega 
Y^*_{L\Lz}(\Omega)
\frac{\cos\theta}{\tilde{\omega}-\cos\theta+i\eta}
Y_{L'\Lz}(\Omega)\,.
\label{eq-res0}
\end{multline}
The Landau parameters $F_{\Lambda\Lambda'}$ can be decomposed into the
direct ($D$) and exchange ($\mathit{Ex}$) contributions from the contact ($c$)
and dipole-dipole ($dd$) interactions, $F_{\Lambda\Lambda'} =
F^{c(D)}_{\Lambda\Lambda'}+F^{c(\mathit{Ex})}_{\Lambda\Lambda'}
+F^{dd(D)}_{\Lambda\Lambda'}+F^{dd(\mathit{Ex})}_{\Lambda\Lambda'}$, and the
corresponding explicit expressions read:
\begin{align}
&F^{c(D)}_{\Lambda\Lambda'}
= \frac{mk_Fg}{2\pi^ 2}(2s+1) 
  \delta_{\Lambda\Lambda'} \delta_{L0}\delta_{S0}\delta_{J0}\delta_{\Jz0}\,,
\label{eq-landau-c-d}\\
&F^{c(\mathit{Ex})}_{\Lambda\Lambda'}
= -\frac{mk_Fg}{2\pi^2} 
  \delta_{\Lambda\Lambda'} \delta_{L0}\delta_{SJ}\,,
\\
&F^{dd(D)}_{\Lambda\Lambda'}
= \frac{2mk_Fc_{dd}}{\pi} s(s+1)(2s+1)\frac{2-3\Jz^2}{9}
\nonumber\\
&\qquad \times\delta_{\Lambda\Lambda'} \delta_{L0}\delta_{S1}\delta_{J1}\,,
\label{eq-dd-direct}
\\
&F^{dd(\mathit{Ex})}_{\Lambda\Lambda'}
= -\frac{5mk_Fc_{dd}}{\pi} s(s+1)(2s+1)
  \delta_{JJ'}\delta_{MM'}
\nonumber\\ 
&\qquad \times 
  \sqrt{(2S+1)(2S'+1)(2L+1)(2L'+1)}
\nonumber\\
&\qquad 
  \times
  (-1)^{S+J} 
  \Biggl[
  (H_L+H_{L'}) 
  \begin{pmatrix} L & L' & 2 \\ 0 & 0 & 0 \end{pmatrix}
  \begin{pmatrix} 1 & 1 & 2 \\ 0 & 0 & 0 \end{pmatrix} 
\nonumber\\
&\qquad 
  +2 (-1)^L \sum_{\ell} (2\ell+1) H_{\ell}
  \begin{pmatrix} L & 1 & \ell \\ 0 & 0 & 0 \end{pmatrix}
  \begin{pmatrix} L' & 1 & \ell \\ 0 & 0 & 0 \end{pmatrix}
\nonumber\\
&\qquad\times
  \begin{Bmatrix} 1 & 1 & 2 \\ L & L' & \ell \end{Bmatrix}
  \Biggr]
  \begin{Bmatrix} L & S & J \\ S' & L' & 2 \end{Bmatrix}
  \begin{Bmatrix} s & s & S \\ s & s & S' \\ 1 & 1 & 2 \end{Bmatrix}
  \,,
\label{eq-landau-dd-ex}
\end{align}
where the standard 3j, 6j and 9j symbols \cite{vmk88} have been used,
and $H_0 = 0$, $H_n = \sum_{p=1}^n 1/p$ are the harmonic numbers.

At this point it seems necessary to add a comment about the direct
term of the DDI in Eq.~(\ref{eq-dd-direct}). As one can see from
Eqs.~(\ref{eq-landau-c-d})$-$(\ref{eq-landau-dd-ex}), the contribution
of the direct term of the DDI is the only one that depends on
$\Jz$. Since the Landau parameters are defined in the limit
$\vek{q}\to 0$, one would expect that there should not be any
preferred direction and therefore no dependence on $\Jz$. In fact, for
$\vek{q} = 0$, one sees immediately that $V^{dd}(\vek{q}=0)=\int d^3r
V_{dd}(\vek{r})=0$, and therefore the direct term was omitted in the
analysis of Ref.~\cite{Fregoso09}. However, the DDI is discontinuous
at $\vek{q} = 0$ because of its long-range nature, and as one can see
from Eq.~(\ref{eq-ddinteraction}), $V^{dd}(\vek{q})$ depends only on
the direction of $\vek{q}$ and therefore does not vanish in the limit
$\vek{q}\to 0$. Since an instability occurs when the energy of a
zero-sound mode (which exists only at small but non-zero $\vek{q}$)
vanishes, we will include the direct term, as it was done in
Ref.~\cite{cwl10}.

Equation~(\ref{eq-RPAeq-J}) can now easily be solved by matrix
inversion:
\begin{equation}
\tilde{\Pi}_{\Lambda\Lambda'}(\tilde{\omega})
=
\sum_{\Lambda_1}
(\mathcal{M}^{-1})_{\Lambda\Lambda_1}
\tilde{\Pi}^0_{\Lambda_1\Lambda'}(\tilde{\omega})\,,
\label{eq-rpa-final}
\end{equation}
where
\begin{equation}
\mathcal{M}_{\Lambda\Lambda'}(\tilde{\omega})
=
\delta_{\Lambda\Lambda'}
-
\sum_{\Lambda_1}
\tilde{\Pi}^0_{\Lambda\Lambda_1}(\tilde{\omega})F_{\Lambda_1\Lambda'}.
\label{eq-matrix}
\end{equation}

The onset of an instability is characterized by a vanishing excitation
energy. We therefore consider the response function in the case
$\tilde{\omega} = 0$. In this case, Eq.~(\ref{eq-res0}) reduces to
$\tilde{\Pi}^0_{\Lambda\Lambda'}(0) = -\delta_{\Lambda\Lambda'}$, so
that $\mathcal{M}_{\Lambda\Lambda'}(0) =
\delta_{\Lambda\Lambda'}+F_{\Lambda\Lambda'}$. Stability requires that
all eigenvalues of $\mathcal{M}(0)$ are positive
\cite{ringschuck,np66,cwl10}. Note that the Landau parameters
$F_{\Lambda\Lambda'}$ and hence the stability matrix $\mathcal{M}$ are
diagonal with respect to the total angular momentum $J$.

Let us now discuss the stability conditions for the special case of
$s=1/2$ atoms. In this case, the total spin $S$ of the excitations can
be $S=0$ or $S=1$.

In the channel $J=0$, there exist two uncoupled modes:
$\Lambda=LSJ\Jz=0000$ and $1100$. The $\Lambda=0000$ mode is
associated with the compressibility and an instability in this channel
means that the system will collapse. The corresponding stability
condition is independent of the DDI since $F^{\rm
  dd(D)}_{0000,0000}=F^{dd(\mathit{Ex})}_{0000,0000}=0$ and reads
\begin{equation}
mk_Fg > -2\pi^2\,.
\label{eq-compress-stability}
\end{equation}

For $\Lambda=1100$, the only contribution comes from the exchange term
of the DDI, since $F^{c(D)}_{1100,1100} = F^{\rm
  c(\mathit{Ex})}_{1100,1100} = F^{dd(D)}_{1100,1100} = 0$. From
Eq.~(\ref{eq-landau-dd-ex}) one obtains $F^{dd(\mathit{Ex})}_{1100,1100} =
mk_Fc_{dd}/(2\pi)$, and since $c_{dd}$ is positive, this mode is
always stable.

In the case of $J=1$, there exist four modes: $\Lambda=101\Jz$,
$111\Jz$, $011\Jz$ and $211\Jz$, the last two being coupled to each
other. For $\Lambda=101\Jz$, no instability can occur because
$F_{1010,1010}=0$.

The response in the $L=0, S=1$ (i.e., $\Lambda = 011\Jz$) channel is
related to the spin susceptibility \cite{jerzak,pethick}, and an
instability in this channel therefore is an instability towards
spontaneous magnetization. Since the $\Lambda = 011\Jz$ and $211\Jz$
channels are coupled, the stability condition is that the matrix
\begin{equation}
\mathcal{M}_{S=1,J=1,\Jz}(0) = 
\begin{pmatrix}1+F_{011\Jz,011\Jz} & F_{011\Jz,211\Jz} \\
F_{011\Jz,211\Jz} & 1+F_{211\Jz,211\Jz}
\end{pmatrix}
\label{eq-spinsusceptibility}
\end{equation}
has positive eigenvalues. This condition depends on both $g$ and
$c_{dd}$. The relevant non-vanishing contributions from the exchange
term of the DDI [Eq.~(\ref{eq-landau-dd-ex})] are 
$F^{\rm  dd(\mathit{Ex})}_{011\Jz,211\Jz}/\sqrt{2} 
= F^{dd(\mathit{Ex})}_{211\Jz,211\Jz} = mk_Fc_{dd}/(12\pi)$. 
Note that
the stability condition depends on $\Jz$ because of the direct term of
the DDI [Eq.~(\ref{eq-dd-direct})]. Physically, the $\Jz = \pm 1$ and
$\Jz = 0$ modes correspond to transverse and longitudinal spin waves,
respectively. For $\Jz = \pm1$, $F^{dd(D)}_{011\Jz,011\Jz}$ is
negative, while for $\Jz = 0$ it is positive. Therefore the $\Jz = \pm
1$ modes become unstable before an instability appears in the $q=0$
case without the direct term of the DDI.

For $\Lambda = 111\Jz$, the only non-vanishing Landau parameter is
$F^{dd(\mathit{Ex})}_{111\Jz,111\Jz}$ which is independent of $\Jz$. The stability 
condition for this channel reads
\begin{equation}
mk_F c_{dd} < 4\pi.
\label{eq-crit111}
\end{equation}

\begin{figure}
\includegraphics[width=30mm]{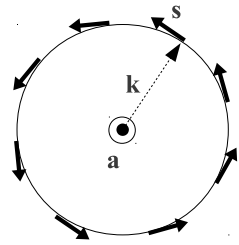}
\caption{\label{fig-external111} Schematic representation of the spin
  texture induced by the perturbation (\ref{eq-external111}) or
  created spontaneously if the system is unstable in the $\Lambda =
  111\Jz$ channel. The circle represents a section though the Fermi
  sphere. The vector $\vek{a}$ points towards the reader.}
\end{figure}

Let us discuss the physical meaning of this instability. The mode
$\Lambda=111\Jz$ can be excited by the following perturbation
\begin{eqnarray}
H'
=
\sum_{\alpha\beta}
\int \frac{d^3k}{(2\pi)^3}\vek{a} \cdot 
(\vek{s}_{\alpha\beta}\times\vek{k})
c^\dagger_{\vek{k}+\vek{q},\alpha}c_{\vek{k},\beta} \, ,
\label{eq-external111}
\end{eqnarray}
where $\vek{a}$ characterizes the amplitude and orientation of the
perturbation. This perturbation tends to turn the spin $\vek{s}$ of an
atom into the direction perpendicular to the momentum $\vek{k}$ and to
the vector $\vek{a}$. As a consequence, the Fermi surfaces of atoms
whose spins point into the favored and unfavored directions split up
and a spin texture in momentum space as shown in
Fig.~\ref{fig-external111}
is created. This is very similar to the so-called Rashba SOC
\cite{rashba60,qz11} which has recently been discussed also in the
context of (two dimensional) systems of ultracold atoms
\cite{Sau11}. If an instability occurs in this channel, this means
that even in the absence of the perturbation (\ref{eq-external111}),
the system will spontaneously choose a direction $\vek{a}$ and create
a spin texture with spins perpendicular to $\vek{a}$ and tangential to
the Fermi surface, as shown in Fig.~\ref{fig-external111}.

\begin{figure}
\includegraphics[width=80mm]{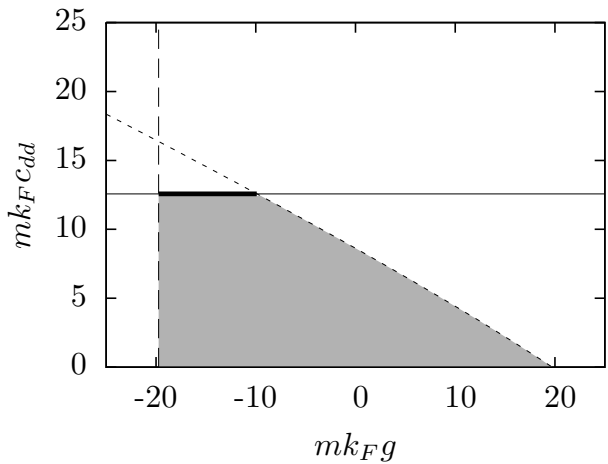}
\caption{\label{fig-instability} The phase diagram for the stability
  of the spin symmetric ground state of the system as function of the
  coupling constants $g$ and $c_{dd}$. The lines indicate the onset of
  different instabilities: collapse (long-dashed vertical line),
  spontaneous magnetisation (short dashes), and spontaneous formation
  of a Rashba-like spin texture (solid horizontal line). The spin
  symmetric ground state is stable in the gray area.}
\end{figure}

The different stability conditions are summarized in
Fig.~\ref{fig-instability}
which shows the stable regions as functions of the coupling constants
$g$ and $c_{dd}$. The long-dashed vertical line indicates the boundary
satisfying Eq.~(\ref{eq-compress-stability}); in the region to the
left of this line, the mean-field theory predicts a collapse. The
short-dashed curve indicates the onset of the instability towards
magnetization obtained from
Eq.~(\ref{eq-spinsusceptibility}). Finally, the region above the
horizontal solid line, obtained from Eq.~(\ref{eq-crit111}), is
unstable with respect to the spontaneous formation of SOC discussed
above. Therefore, the stable region of the spin symmetric ground state
is the gray area. The region where one can expect to find the
spontaneous formation of SOC is situated above the thick part of the
solid line. For completeness, it should be mentioned that the modes
with $J\ge 2$ do not change the phase diagram: For $J\ge 2$, the
contact interaction does not contribute and the critical values for
$c_{dd}$ are always larger than that given by Eq.~(\ref{eq-crit111}).

Notice that the present stability analysis in the framework of RPA is
only able to detect second-order phase transitions. To investigate the
stability against first-order phase transition, one would have to use
other methods, e.g., a variational ansatz as in Ref.~\cite{ff09}.

\begin{figure}
\includegraphics[width=80mm]{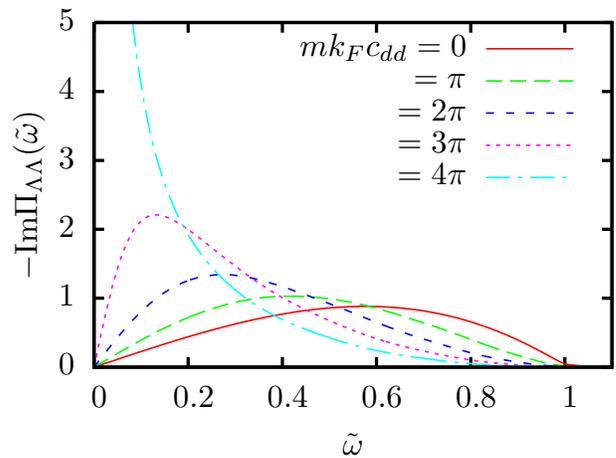}
\caption{\label{fig-1110mode}
The imaginary part of the response function (\ref{eq-rpa-final})
for $\Lambda=\Lambda'=1110$ for various values of $c_{dd}$.}
\end{figure}

So far we have only considered the response at $\tilde{\omega}=0$. But
using the RPA, we can also calculate the energy spectrum of the
corresponding zero-sound modes. This is somewhat more difficult
because at $\tilde{\omega}\neq 0$ the response function
$\Pi^0_{\Lambda\Lambda'}$ is no longer diagonal with respect to $J$
(although it remains diagonal with respect to $\Jz$) and, therefore,
an infinite number of modes are coupled among one another in the
matrix $\mathcal{M}(\tilde{\omega})$ of Eq.~(\ref{eq-matrix}). To
calculate numerically the response function, it is necessary to
truncate the matrix at some value $J_{\mathit{max}}$. However, we
found that convergence is practically reached at $J_{\mathit{max}}
\sim 10$. As an example, we display in Fig.~\ref{fig-1110mode}
the imaginary part of the response function in the channel $\Lambda =
1110$ for different values of the coupling constant $c_{dd}$. Since
the Landau parameter $F_{1110,1110}$ is always negative, there is no
zero-sound mode but only a broad particle-hole continuum. As one
approaches the instability at $mk_Fc_{dd} = 4\pi$, the response gets
more and more enhanced at low energy and at the instability it
diverges at $\tilde{\omega}=0$.

To summarize, we have applied mean-field and RPA theory to an
unpolarized dipolar Fermi gas in the spin symmetric ground state. We
have discussed the stability conditions in the special case of
spin-$1/2$ atoms. In addition to the known collapse and spontaneous
magnetization instabilities, we found that for certain values of $g$
and $c_{dd}$ the system gets unstable towards a phase with SOC where a
Rashba-like spin texture around a spontaneously chosen axis in
momentum space is formed.

An important subject for future studies will be to check whether the
SOC phase survives also beyond the mean-field approximation. For
instance, in reality the collapse predicted by the Hartree-Fock
approach in the case of a strongly attractive contact interaction does
never occur and the Fermi gas stays stable even in the unitary limit
\cite{Heiselberg}. The stabilizing effect of short-range correlations
was also discussed in the context of the instability towards
spontaneous magnetization in the case of a strongly repulsive contact
interaction (without DDI)
\cite{Zhai}. In nuclear physics, short-range
correlations due to the tensor force, which is similar to the DDI, are
known to be important \cite{Feldmeier}.

In addition, it is interesting to see what happens if the gas is
trapped and not uniform. It will also be important to study not only
the onset of the instability, but also the SOC phase itself and the
competition between the different phases. Another interesting question
is whether the SOC phase persists in (quasi-) two dimensional
systems. In this case it would be possible to study, e.g., the spin
Hall effect with cold atoms.

We thank D. Basko for interesting discussions and suggestions.

\end{document}